# A Bayesian group sequential schema for ordinal endpoints

Chengxue Zhong[a], Hongyu Miao[b,c], Haitao Pan[d,*],

[a] *Department of Biostatistics and Data Science, University of Texas Health Science Center at Houston, TX, USA*

[b] *College of Nursing, Florida State University, Tallahassee, FL, USA*
[c] *Department of Statistics, Florida State University, Tallahassee, FL, USA*
[d] *Department of Biostatistics, St. Jude Children's Research Hospital, Memphis, TN, USA*

[*]Correspondence: haitao.pan@stjude.org

**Abstract**

Ordinal endpoints are useful in clinical studies. For example, many clinical trials for hospitalized patients with COVID-19 used an ordinal scale recommended by the World Health Organization. Despite their important role in clinical studies, design methods to assess ordinal endpoints are limited; in practice, a dichotomized approach is often used. Here we introduce a Bayesian group sequential schema to assess ordinal endpoints. The proposed designs include a proportional-odds (PO) model, a nonproportional-odds (NPO) model, and a PO/NPO-switch model to handle various scenarios. Extensive simulations are conducted to demonstrate desirable performance. Finally, we developed the R package *BayesOrdDesign* to implement the proposed methods.

## 1. Introduction

In clinical studies, the familiar primary endpoints include continuous, binary, and time-to-event points. However, the ordinal endpoint, which is less frequently regarded as the primary endpoint, has recently gained attention for its use in drug-development studies in the current pandemic era. For example, among 49 Phase III randomized trials for COVID-19 that were registered by April 2020, the most common primary endpoint was the ordinal endpoint (Sakamaki et al., 2021). The World Health Organization suggests using the ordinal scale from 0 (no clinical or virological evidence of infection) to 8 (death) as the primary endpoint (Synopsis, 2020). The international ATTACC trial, which is an open-label, randomized, controlled trial, evaluated whether therapeutic-dose anticoagulation with low-molecular-weight heparin or unfractionated heparin prevented mechanical ventilation and/or death in hospitalized patients with COVID-19, compared to usual care (Houston et al., 2020). The primary outcome of that study was an ordinal endpoint with three possible outcomes based on the worst status of each patient through Day 30: no requirement for invasive mechanical ventilation, invasive mechanical ventilation, or death. The ordinal endpoint was chosen because it increases statistical power compared with a binary endpoint and is justified based on clinical importance.

Ordinal endpoints are also useful in oncology studies. For instance, Phase II trials frequently compare doses of novel agents or different combinations of therapy to choose a candidate treatment for the Phase III trial. Although the binary endpoint (response vs no response) is often considered, it can be more informative to use an ordered categorical outcome comprising three or more states: complete response, partial response, stable disease, progression, or death. In some fields, the ordinal endpoint has been accepted into routine clinical practice. For instance, the Breast Imaging Reporting and Data System considers mammography assessments as ordered categories, numbered from 0 to 6 (Liberman & Menell, 2002). Using these ordinal outcomes, clinicians can consistently report their findings on mammograms, which benefits subsequent communications and follow-up interventions. Furthermore, the ordinal endpoint provides higher statistical power than does dichotomization of endpoints, as seen in several real data sets from studies of various diseases (e.g., stroke, influenza, traumatic brain injury, cirrhosis, and respiratory syndrome) (D'Amico et al., 2020; Peterson et al., 2019; Roozenbeek et al., 2011; Whitehead & Horby, 2017; Whitehead & Jaki, 2009).

Although ordinal endpoints play important roles in clinical studies, design methods for the ordinal endpoints are still limited. Whitehead (1993) derived a sample-size formula for ordinal endpoints by comparing a treatment regimen with a control in a superiority-design setting (Whitehead, 1993). A proportional-odds (PO) model was assumed in the derivation. Whitehead and Jaki (2008) proposed one- and two-stage designs based on an approach similar to that of Dunnett but for the ordinal endpoint (Whitehead & Jaki, 2009). Whitehead and Horby (2017) proposed a generic ordinal sequential trial design (using a large-sample theory) for a randomized clinical trial comparing an experimental treatment with standard care for Middle East respiratory syndrome coronavirus (Whitehead & Horby, 2017). The trial was monitored using a series of as many as 20 interim analyses, equally spaced to include responses in newly accrued patients. In a blog, Harrell and Lindsell (2021) proposed a Bayesian adaptive sequential design for the ordinal endpoint with R code exhibitions; they also extended the longitudinal ordinal outcomes (Harrell & Lindsell, 2021). It should be noted that all the above-mentioned methods assumed PO models for the ordinal endpoints (McCullagh, 1980).

We used the conceptual example presented in Table 1 to describe the PO model and related hypothesis-testing problems. With an ordered categorical response, essential data on treatment efficacy, drawn from a total of $n$ patients, can be summarized as shown in Table 1, where the first subscript 1 and 2 denote the control and experimental groups, respectively. The possible categories of outcome are labeled $1, \ldots, C$, with the smallest value (1) being the most desirable outcome, and the largest value ($C$) being the worst.

Let $p_{ic}$ denote the probability that an individual $i$ receiving the experimental treatment has a response in category $c$, and $P_{ic}$ represents the probability of the outcome $c$ or better,

$$P_{ic} = p_{i1} + \cdots + p_{ic}, c = 1, \ldots, C$$

($P_{iC} = 1$).

Let $q_{jc}$ and $Q_{jc}$ denote similar definitions for the control group. The parameter log-odds-ratio $\Delta_c$ is then defined as follows:

$$\Delta_c = log\left(\frac{P_{ic}(1 - Q_{jc})}{Q_{jc}(1 - P_{ic})}\right).$$

Thus, $\Delta_c$ measures the tendency of patients in the experimental arm of the study to be more likely to fall into category $c$ or better. It is a measure of the efficacy of the experimental treatment over the control, and a positive $\Delta_c$ indicates a promising effect of the experimental treatment in patients. Under the PO assumption, we assume $\Delta_1 = \cdots = \Delta_{C-1}$ and denote their common value by $\Delta$. Under this assumption, the above design using the PO model with the ordinal outcome is most powerful to detect a treatment effect $\Delta$ and have a tighter confidence interval than other ordinal methods or approaches that dichotomize the ordinal scale (DeSantis et al., 2014). However, when the PO assumption does not hold, the power may be compromised (Choi, 2016).

Several alternatives to the PO model have been proposed, such as the sliding dichotomy based on a baseline covariate, assumption-free tests, and the randomization test (Diener et al., 2008; Murray et al., 2005; Howard et al., 2012). Although these approaches address the nonproportionality of the odds, they fail to exploit all the information provided by the ordinal scale of the outcome measure, which is not optimal in all applications (Bath et al., 2007; Desantis et al., 2014). To the best of our knowledge, the literature includes one Bayesian design (Murray et al., 2018) in which the comparative-testing criterion is based on utilities for the levels of the ordinal outcome and the PO model and a nonproportional odds (NPO) model, with a hierarchical prior that shrinks toward the PO model used. For the NPO model, Guo and Yuan (2017) used the dispersed cumulative probit model of McCullagh (1980) for personalized dose finding in a Phase I/II study of molecularly targeted agents (Guo & Yuan, 2017). However, in practice, how one chooses between PO and NPO models and how this uncertainty is incorporated into design planning are important questions that still lack answers.

In this paper, we explore comprehensively the Bayesian PO models and NPO models and propose a novel PO/NPO-switch model in a two-arm group sequential design (GSD) setting with the ordinal endpoint. The first design is based on the PO cumulative logistic-regression model, which is similar to that of Harrell and Lindsell (2021), in which the log-odds ratios for all categories are assumed to be equal, e.g., $\Delta = \Delta_1 = \cdots = \Delta_{k-1}$. However, we provide a more formal Bayesian design with a sample-size determination procedure based on numerical simulations. For the NPO model–based design, we used a cumulative logistic location-shift model to capture the heterogeneities of the log-odds ratios among different outcome categories;

thus, not all $\Delta_1, \cdots, \Delta_{k-1}$ are equal. Due to this complexity, the efficacy summary measure for comparing the experimental groups to the control group was not easily evaluated using a single log-odds ratio. Instead, we adopted a utility-based method (Murray et al.2018) to propose a Bayesian GSD.

It is often challenging to know which model, PO or NPO, is the appropriate at the start of a trial. However, if information can be collected from an initial running phase, it may help us select an appropriate model. To that end, we propose a novel, robust two-stage PO/NPO-switch model design schema for ordinal endpoints. A reversible-jump Markov Chain Monte Carlo (RJMCMC) algorithm is developed to allow a data-driven switch between PO and NPO structures after the data from the running phase (first stage) are observed and a procedure for sample-size re-estimation is also introduced. Of note, the choice of Bayesian model in cumulative-link ordinal-regression models has also been discussed by McKinley et al. (2015) but not in the context of trial design.

The remainder of the paper is organized as follows: Section 2 first introduces the proposed PO and NPO models and corresponding design frameworks. Then, it introduces the PO/NPO-switch model–based schema and the RJMCMC algorithm. Section 3 examines the operating characteristics by simulations. Section 4 illustrates how to use the R package *BayesOrdDesign* to implement the proposed methods. We conclude with a brief discussion in Section 5.

## 2. Methods

### 2.1. Proportional odds model–based design

Let $X = (x_1, x_2, \ldots, x_n)'$ and $Y = (y_1, y_2, \ldots, y_n)'$ be the observed ordinal outcomes for two treatment groups that originated from the unobserved latent variables $X^*$ and $Y^*$. To be specific, let $X^*$ and $Y^*$ be the unobserved latent outcomes for the control and experimental groups, respectively, where $X^* = (x_1^*, x_2^*, \ldots, x_n^*)'$, and $Y^* = (y_1^*, y_2^*, \ldots, y_n^*)'$. Assume that the latent variable follows a logistic distribution with a common standard deviation σ, then the corresponding linear-regression models are as follows:

$$x_i^* = \mu + \varepsilon_i, \quad \varepsilon_i \sim Logistic(0, \sigma^2), i = 1, \ldots, n, \quad (1)$$

$$y_j^* = \mu + \Delta + \eta_j, \ \eta_j \sim Logistic(0, \sigma^2), j = 1, \ldots, n, \quad (2),$$

where the sample sizes of the control group and the treatment group are assumed to be the same; thus, $I = J = n$, (e.g., 1:1 equal allocation). For simplicity, we replace $I$ and $J$ with $n$ in the rest of the discussion. Assume that the observed ordinal variables $X = (x_1, x_2, \ldots, x_n)'$ and $Y = (y_1, y_2, \ldots, y_n)'$ originate from the, unobserved variables $X^*$ and $Y^*$. The latent thresholds $\gamma_c$ ($1 < c < C - 1$) and $\theta_c$ ($1 < c < C - 1$) partition the value of $X^*$ and $Y^*$ into the $C$ ordinal categories of $X$ and $Y$. Formally, the correspondence between the latent variables and the observed outcomes are described as shown below:

$$x_i = c, \text{ if } \gamma_{c-1} < x_i^* < \gamma_c,$$

$$y_j = c, \text{ if } \theta_{c-1} < y_j^* < \theta_c,$$

$$c = 1, \ldots, C - 1,$$

where the boundaries $\gamma_c$ and $\theta_c$ are unknown, and $-\infty = \gamma_0 < \gamma_1 < \cdots < \gamma_{C-1} < \gamma_C = \infty$.

We let $\boldsymbol{\gamma} = (\gamma_1, \ldots, \gamma_{C-1})$ and $\boldsymbol{\theta} = (\theta_1, \cdots, \theta_{C-1})$ denote the vector of thresholds. For the control group, assume that $\varepsilon$ has a logistic-distribution function $F(\cdot)$ whose range lies in $(-\infty, \infty)$:

$$P(\varepsilon \leq z) = F(z). \qquad (3)$$

Combine Eqns. (1), (2), and (3) to obtain

$$P(x_i \leq c|\mu) = P(x_i^* \leq \gamma_c|\mu) = P(\varepsilon_i + \mu \leq \gamma_c) = P(\varepsilon_i \leq \gamma_c - \mu) = F(\gamma_c - \mu),$$

where $P(x_i \leq c|\mu)$ is denoted as $P_{ic}$ and refers to the cumulative probability for the control group. The probability $P(x_i = c|\mu)$, which is of primary interest, can be computed from

$$p_{ic} = P(x_i = c|\mu) = F(\gamma_c - \mu) - F(\gamma_{c-1} - \mu).$$

Similarly, for unobserved ordinal variables in the experimental group, the probability $P(y_i = c|\mu, \Delta)$ and the cumulative probability $P(y_i \leq c|\mu, \Delta)$ for the treatment group, denoted as $Q_{ic}$, can be derived from

$$q_{ic} = P(y_j = c) = F(\theta_c - \mu - \Delta) - F(\theta_{c-1} - \mu - \Delta).$$

If we assume that $\gamma_c = \theta_c$ for $c = 1, \ldots, C - 1$, then

$$\log\left(\frac{P_{ic}(1-Q_{jc})}{Q_{jc}(1-P_{ic})}\right) = \Delta, \qquad (4)$$

where $\Delta$ represents the log OR between the two groups, and it implies that the treatment effect is constant, regardless of the outcome scale. Therefore, if the PO assumption holds, we will test the following hypothesis:

$$H_0: \Delta = 0 \ vs. \ H_a: \Delta < 0.$$

The null hypothesis $H_0$ implies no difference between the two groups (i.e., this is equivalent to all log-odds ratios being equal), and the alternative, $H_a$, suggests an effective treatment (i.e., a higher value in the category indicates a less desirable outcome). The model above can be evaluated in a Bayesian framework by using Markov Chain Monte Carlo (MCMC). To construct a Bayesian GSD, we can use the following criterion to conduct the interim monitoring and make the final decision:

$$\pi = \Pr(\Delta < 0|\mathcal{D}). \qquad (5)$$

Here, D is the accumulative information at the interim or final stage, and a high value of $\pi$ means that the experimental group is better than the control. The proposed Bayesian design is as described below, if one interim look and one final look are planned. Herein, $n$ patients are recruited into two groups, respectively, with a 1:1 allocation ratio at the interim stage.

(a) Enroll $2n$ patients and equally randomize them into treatment and control groups during the first stage.

(b) Given the first-stage data, $\mathcal{D}_{trt} = (y_{trt,1}, \ldots, y_{trt,n})$ and $\mathcal{D}_{ctr} = (y_{ctr,1}, \ldots, y_{ctr,n})$. If $\pi < C_f$, then terminate the trial early due to futility, where $C_f$ is a small value (e.g., 0.2) of the cutoff point for futility stopping.

(c) Otherwise, at the second stage continue to enroll an additional $2n$ patients and equally randomize them to the treatment or control groups.

(d) Once the maximum sample size is reached, calculate $\pi$ based on pooled data from the two stages. If $\pi > C_s$, where $C_s$ is a large value (e.g., 0.9) of the cutoff point for superiority, then conclude that the treatment is superior; otherwise, conclude that the treatment is ineffective.

To guarantee desirable frequentist-operating characteristics in this design, futility and superiority cutoff points $C_f$ and $C_s$ need to be calibrated to achieve targeted type I (under $H_0$) and type II ($H_a$) error rates by numerical searching via simulations.

## 2.2. Nonproportional odds model–based design

To describe NPO ratios among different categories without the assumption of equality between $\gamma_c$ and $\theta_c$ for $c = 1, \ldots, C-1$, we assume the following generic model:

$$\log\left(\frac{P_{ic}(1-Q_{jc})}{Q_{jc}(1-P_{ic})}\right) = \Delta + \gamma_c - \theta_c = \Delta + \Delta_c = \Delta'_c. \quad (6)$$

$$\Delta_c \sim Normal(\widehat{\Delta}_c, \hat{\sigma}_{\Delta_c}^2).$$

By taking Bayesian approaches, prior distributions for μ, γ, and θ are specified (in the Appendix). In the model (6), prior distribution for $\Delta_c$, such as mean and variance, can be estimated via the information available at the interim stage and by using the R package *ordinal*. (Christensen RHB, 2019) However, comparing treatments based on an NPO model is more complex than that based on the PO model because even if $\Delta_c$ for each outcome scale is known, assessing the efficacy of treatment is not straightforward. For example, when comparing two groups based on a three-category outcome (e.g., Good, Intermediate, and Poor), assuming the experimental group has true proportions of (0.3, 0.6, 0.1) being in Good, Intermediate, and Poor categories, and the control group has true proportions of (0.4, 0.2, 0.4) for each of these categories. We see that, with respect to the "Good" category, the control group (0.4) is better than the experimental group (0.3), and the experimental Intermediate category (0.6) is superior to that in the control group (0.2) by comparing the proportions of responses. Therefore, whether the experimental group is superior to the control is unclear. To that end, we adopt the utility-based method proposed by Murray et al. (Murray et al., 2016) to develop a criterion for determining how to compare two groups in this context. By this method, numerical values for utility (defined as the utility scores) for all outcome categories need to be specified. To compare two groups, the utility scores can be interpreted in a relative scale. Therefore, these values can be conveniently elicited by investigators (see examples below). Given the utility scores, we can then compare the two groups by using a mean utility as defined below:

$$\overline{U}(\pi(trt)) = \sum_{c=1}^{C} U_c \times \pi_c(trt), \quad (7)$$

where $\pi_c(trt)$ represents the probability of each category from 1 to $C$ from the experimental group, and the utility function $U_c$ denotes the corresponding elicited utility score for each outcome category. Similarly, we let $\pi_c(ctr)$ denote the probability of each category for the control group and define $\overline{U}(\pi(crt))$ as the mean utility for the control group. For example, when the utility scores are $U_1 = 100, U_2 = 60,$ and $U_3 = 0$ and the true probabilities are $\pi_1(trt) = 0.3, \pi_2(trt) = 0.6, \pi_3(trt) = 0.1, \pi_1(ctr) = 0.4, \pi_2(ctr) = 0.2, \pi_3(ctr) = 0.4$ as aforementioned, the mean utilities are 66 and 52 for the experimental group and control group, respectively. Given the mean utilities, the experimental group is obviously clinically superior to the control group.

In our context with the six categories for the outcomes, category 1 is the most desirable outcome, and category 6 is the least desirable. Therefore, to elicit the utility scores, we can first set $U_1 = 100$ and $U_6 = 0$. Utility numerical values for the rest of the categories can be specified by physicians based on their clinical rationale. For example, for categories 2 to 5, the utility scores can be set as

$$U_2 = 80, U_3 = 65, U_4 = 25, \text{ and } U_5 = 10.$$

Once the utility scores have been set, the Bayesian model can be used to estimate $\pi_c(trt)$ and $\pi_c(crt)$ and then the mean utility scores of $\overline{U}(\pi(trt))$ and $\overline{U}(\pi(crt))$ for the experimental and control group, respectively. For this specific design purpose, we can use the following estimated $\pi_U$ as a cutoff to determine whether the experimental group is superior to the control:

$$\pi_U = Pr\{\overline{U}(\pi(trt; \alpha)) > \overline{U}(\pi(crt; \alpha))|\mathcal{D}\}. \quad (8)$$

Here, $\mathcal{D}$ represents the aggregated data from both groups, and $\alpha$ denotes the unknown parameter sets.

With the proposed comparative criterion, the corresponding Bayesian design is similar to the aforementioned PO model design. After the first-stage data are obtained, $\pi_U$ can be calculated. The trial is terminated early if $\pi_U < C_f$. Otherwise, patient enrollment continues, and $\pi_U$ is evaluated based on all the data. If $\pi_U > C_s$, we conclude that treatment is superior. For the

NPO model design, $C_f$ and $C_s$ should be calibrated to control the type I and II error rates through simulations.

Two Bayesian designs for the ordinal endpoint, the PO model– and NPO model–based designs, have been introduced so far. We see that the PO model is a special case of the NPO model, i.e., the NPO model can reduce to the PO model if the same set of utility scores are used and the proportions for all outcome categories are assumed to be equal. However, the PO and NPO are distinct models with different sets of parameters. By simulations, we found that the two designs have different results, even given the same setting and sample size. For example, to control the type I error rate at 5%, given the PO assumption and odds ratio (OR) of 1.8 under $H_a$ to be detected by 1,000 simulations, a total sample size of 300 is required to achieve 84.2% power; the NPO model–based design has a lesser power of 82.1% with the sample size of 300. This difference may be caused by the NPO model having more parameters than the PO model.

Similar results can be found if the true OR is $(1.9, 1.8, 1.8, 1.6, 1.4)$, by using the elicited utility scores of $(U_1 = 100, U_2 = 80, U_3 = 65, U_4 = 25, U_5 = 10, U_6 = 0)$, the effect size detected is 9.05, which is close to the mean utility score of OR = 1.8, that is 9.20 in terms of the utility score. In this setting, a total sample size of 200 based on the NPO model–based design has 98.9% power, and the PO model has 64.3% power with the total sample size of 200. From this result, we notice that the difference in the powers between the PO and the NPO models is not negligible. This difference can be explained, in part, by the effect size depending on not only the alternative OR of each outcome category but also the elicited utility scores. Specifying different utility scores may result in different effect sizes, which in turn, need a different sample size; even the alternative ORs at each outcome category are the same.

Some conclusions can be made from the above discussions: (i) we cannot simply regard the PO model as a special case of the NPO model when designing a study; (ii) a design based on the NPO model depends on not only the alternative OR at each outcome category but also how to elicit the utility scores; (iii) even under the "same" setting, if the model is incorrectly assumed, a different sample size may be required, and the targeted power may not be achieved. However, it is challenging to prespecify which model, PO or NPO, should be used before the onset of a trial. Therefore, in the next section, we propose a PO/NPO-switch model–based design.

### 2.3. PO/NPO-switch model–based design

The PO model– and NPO model–based designs have been introduced. However, before a trial is initiated, there are uncertainties about which model is more appropriate. Thus, a reasonable approach would be to first have a running phase (first stage) of the trial to gain some information that will help us select the more appropriate model. We can then enroll patients by using the design determined during the first stage and make final decisions at the end of the second stage. For the goal of "selection" in the first stage, we propose a method that uses the RJMCMC algorithm. Because the model may be switched from the PO to the NPO (or vice versa) after the first stage, we call this method the PO/NPO-switch model–based design.

As outlined by Green (Green, 1995), a key step in the RJMCMC is to build bijective functions describing the relations between parameters from different models. When the dimensions of two models differ, supplemental parameters are introduced to ensure consistency in the parameter number. For instance, if model $M_1$ has 5 parameters and model $M_2$ has only 1 parameter, then to map $M_2$ to $M_1$, a bijective function $g$ is required with a supplemental variable vector $\boldsymbol{\mu_2}$, where $\boldsymbol{\mu_2}$ is 4-dimensional but is not connected with $M_2$, as it serves as a place holder only for $M_2$. Thus, to build connections among various models, we need to specify $\binom{K}{2}$ bijective functions, where $K$ is the number of candidate models.

When $K$ is large, the number of bijective functions could result in a computational challenge. Therefore, to simplify the RJMCMC, Barker & Link (Barker & Link, 2013) proposed a universal parameter $\boldsymbol{\psi}$ called the "palette", from which all parameters of different models can be calculated. Specifically, parameter vector $\boldsymbol{\xi}$ can be converted through a known mapping function $g(\boldsymbol{\psi}) = \boldsymbol{\xi} = (\boldsymbol{\beta}, \boldsymbol{u})$. With the palette in hand, the number of bijective functions is reduced from $\binom{K}{2}$ to $2K$, as there is no need to develop bijective functions for each pair of models. Crucially, this implementation is a post-processing algorithm, which means that samples from model-specific posteriors $p(\boldsymbol{\xi}|data, Model)$ can be used directly, once the candidate models are fitted. In our context, we consider the two models:

$$M_1: NPO\ model, with\ parameter\ \boldsymbol{\beta_1}$$

$$M_2: PO\ model, with\ parameter\ \boldsymbol{\beta_2}.$$

We construct a 5-dimensional palette by assuming 6-category ordinal endpoints, $\boldsymbol{\psi} = (\psi_1, \ldots, \psi_5)$. In the NPO model, parameter vectors $\boldsymbol{\beta_1} = (\Delta_1, \ldots, \Delta_5)$ are linked to $\boldsymbol{\psi}$; in the PO model, $\Delta$ is associated with the weighted average $\bar{\psi} = \frac{1}{5}\sum_{c=1}^{5}\psi_c$ and a supplemental vector $\boldsymbol{u}=(\psi_2, \ldots, \psi_5)$. Thus, bijective functions $g_1(\boldsymbol{\psi}) = \boldsymbol{\psi} = \boldsymbol{\beta_1}$ and $g_2(\boldsymbol{\psi})$ are defined as follows:

$$g_2(\psi_1, \ldots, \psi_5) = (\bar{\psi},\ \bar{\psi} - 2\psi_2, \ldots, \bar{\psi} - 2\psi_5)$$

$$= (\Delta, u_1, \ldots, u_4),$$

and

$$g_2^{-1}(\Delta, u_1, \ldots, u_4) = (\Delta + \sum_{c=1}^{4} u_c, \Delta - u_1, \ldots, \Delta - u_4)$$

$$= (\psi_1, \ldots, \psi_5).$$

Our RJMCMC sampler for the PO/NPO-switch model design proceeds as follows:

(a) Given the current value of $M$, sample $\boldsymbol{\psi}$:

- Under $M_1$, we sample $\Delta_i$ from $[\Delta_i|M_1, \mathcal{D}]$, for $i = 1, \ldots, 5$. We then compute $\boldsymbol{\psi} = (\Delta_1, \ldots, \Delta_5)'$.
- Under $M_2$, we sample $\Delta$ from $[\Delta|M_2, \mathcal{D}]$ and $u$ from $[u|M_2]$. We then compute $\boldsymbol{\psi} = g_2^{-1}(\Delta_1, u_1, u_4)'$.

(b) Given the current value of $\boldsymbol{\psi}$, sample $M$. We use Eqn. (5) to compute full conditional model probabilities. First, we calculate model-specific parameters ($\boldsymbol{\beta_k}$) from $\boldsymbol{\psi}$. The likelihood $[\mathcal{D}|\boldsymbol{\psi}, M_k]$ is simply $[\mathcal{D}|\boldsymbol{\beta_k}, M_k]$; the prior $[\boldsymbol{\psi}|M_k]$ is computed via Eqn. (6).

$$\Pr(M_k|\cdot) = \frac{[\mathcal{D}|\boldsymbol{\psi}, M_k][\boldsymbol{\psi}|M_k][M_k]}{\sum_j [\mathcal{D}|\boldsymbol{\psi}, M_j][\boldsymbol{\psi}|M_j][M_j]}, \quad k = 1, 2; \quad (5)$$

$$[\boldsymbol{\psi}|M_k] = f_k(g_k(\boldsymbol{\psi}))\left|\frac{\partial g_k(\boldsymbol{\psi})}{\partial \boldsymbol{\psi}}\right| \quad (6)$$

With the specifications of log-likelihood $[\mathcal{D}|\boldsymbol{\psi}, M_k]$ and prior distributions $[\boldsymbol{\psi}|M_k]$, we sample $M_1$ with the probability $Pr(M_1|\cdot)$; otherwise, we sample $M_2$ with the probability

$Pr(M_2|)$. R package rjmcmc (Gelling et al., 2018) was used for the model selection. In the Appendix, we show detailed specification of the likelihood and prior distributions. Procedures of the proposed PO/NPO-switch model–based design are shown below and in Figure 1.

Given a targeted type-I error rate, power, and the expected effect size (e.g., log-odds ratio for the PO model–based design and difference of mean utility scores of two groups for the NPO model–based design):

(a) Compare the estimated overall sample size $N_{po}$ from the PO model structure and $N_{npo}$ from the NPO model structure. Select the larger value $N_{max} \in (N_{po,1}, N_{npo,1})$ as the sample size for the first stage, where $N_{po,1}$ and $N_{npo,1}$ denote the sample sizes of the first stage from the PO and NPO model–based designs, respectively.

(b) Given the first-stage data $\mathcal{D} = (y_{trt,1}, \dots, y_{trt,N_{max}}, y_{ctr,1}, \dots, y_{ctr,N_{max}})$, apply the RJMCMC approach to select the best-fit model M.

(c) Given the model M from step (b) and data from the first stage and using the criterion in Eqns. (5) or (8) to determine whether $\pi < C_f$, terminate the trial early and conclude that the treatment efficacy is not promising enough, where $C_f$ is the threshold for the futility-stopping rule derived either from the PO- or NPO-based design and step (b).

(d) Otherwise, continue the trial to enroll $N_M$ patients in treatment and control groups for the second stage, where $N_M$ could be either $N_{po,2}$ or $n_{npo,2}$; here, $N_{po,2} = N_{po} - N_{po,1}$ and $N_{npo,2} = N_{npo} - N_{npo,1}$ refers to the rest of the sample sizes for the PO model– and NPO model–based designs, respectively. Once the $N_M$ sample size is reached, calculate $\pi$ based on the pooled data from the two stages. If $\pi > C_s$, conclude that the treatment is effective; otherwise, the treatment is not effective.

Notes for the above procedures:

For step (a), we will initially ask the investigator to provide two sets of alternative hypotheses (one for PO and another for NPO). This is feasible because the investigators have good clinical rationale for how they set up the hypotheses, e.g., the expected effect size represented as either a log-OR (for the PO model) or expected log-ORs for outcome categories, which can be translated into the difference of mean utility scores (for the NPO model) and will be used for sample-size determinations. For the PO and NPO, the expected effect size provided by the investigators may result in different sample sizes to claim a successful treatment group. Thus, either sample size (i.e., PO is larger or NPO is larger) is possible. We choose the larger sample size as the initial sample size for the design (shown as $N_{max} \in (N_{po,1}, N_{npo,1})$) in step (a) to let the design have adequate power, though this strategy might be conservative.

For steps (b) and (c), we let the design be a two-stage trial: after the first stage and based on the PO/NPO-switching algorithm, we decide which model would be a reasonable choice, as evidenced by the first-stage data.

For step (d), there are four scenarios: (i) If the initial sample size is based on the PO model, which is now recommended, then the second stage continues as planned. For example, $N_{max} = N_{po,1}$ and the sample size for the second stage is $N_M = N_{po,2} = N_{po} - N_{po,1} = N_{po,2} = N_{po} - N_{max}$. (ii) If the initial sample size is based on the PO model, then the NPO model requires fewer sample sizes than the PO model and now recommends the NPO model. Therefore, we have the opportunity to save the sample size during the second stage. To be technically specific, we can use the first-stage data to predict how many sample sizes would be used for the second stage based on the NPO model, and we can apply the sample size re-estimation strategy in this situation. However, for simplicity in this paper we have used the sample size based on the PO model in the second stage. (iii) If the initial sample size is based on the NPO and now recommends the NPO model, then the second stage continues as planned. (iv) If the initial sample size is based on the NPO model (it means that the PO model requires fewer sample sizes than the NPO model), and it now recommends the PO model after the first stage. In this situation, we have the opportunity to save the sample size during the second stage. Similar to (ii), we can use the first-stage data to predict how many sample sizes would be used for the second stage based on the PO model, and we can apply the sample size re-estimation strategy. For simplicity, in this paper we have used the large sample size based on the NPO model in the second stage. Of note, the NPO model commonly requires more sample sizes than the PO model, though the required sample size of the NPO model depends on the elicited utility scores.

Like the previously proposed PO model– and NPO model–based designs, to achieve targeted frequentist operating characteristics of the design, the futility and superiority cutoff points $C_f$ and $C_s$ need to be calibrated to control type I and II errors. A grid search for cutoff points is used via simulations.

## 3. Simulations

### 3.1. Simulation setting

In the simulations, we implemented a two-stage, two-arm setting, with the ordinal endpoint for the three above-proposed designs, where the patients are recruited with an equal randomization. For the PO model–based design, assuming the OR for ordinal outcomes is 1.8 under the alternative hypothesis, 341 patients are needed to achieve 80% power, given the type-I error rate controlled at 5%. To assign patients equally during the first stage, we enroll 100 patients for each group and conduct the interim analysis. Then, we terminate the trial if the futility criteria are met based on the interim analysis; otherwise, we continue to enroll another 100 patients for each group and conduct the final analyses. For both the NPO model–based and PO/NPO-switch model–based designs, we enroll 100 patients for each group at the first and second stages. Thus, the total sample size of patients is 400.

For the control group, $q_{jc} = (0.58, 0.05, 0.17, 0.03, 0.04, 0.13)$ is assumed to be proportions of occurrence in each outcome category. The distribution of outcome categories $p_{ic}$ for a treatment group of 21 scenarios is shown in Table 2. In addition, utility scores for the six outcomes are assumed to be elicited by investigators as $U_c = (100, 80, 65, 25, 10, 0)$. Corresponding numeric mean utilities of $p_{ic}$ and $p_{jc}$, for control and treatment groups under each scenario, can be found in columns $\bar{U}_c, \bar{U}_t$, and $\bar{U}_t - \bar{U}_c$ in Table 2. For example, the mean utility score $(\bar{U}_c)$ for the control is 74.20 (calculated by $q_{jc} \times U_c$) and for the treatment group $(\bar{U}_t)$ is 75.88 for Scenario 1; thus, and the difference in mean utility scores $\bar{U}_t - \bar{U}_c$ is 1.68.

[Table 2]

The type-I error rate is controlled at 5%, and the futility and superiority thresholds $C_f$ and $C_s$ are calibrated via simulations. The calibrated thresholds for the three designs are listed in Table 3.

[Table 3]

Under the alternative hypothesis, various scenarios are investigated via simulations using the three proposed designs to explore the properties of the proposed approaches (Table 2). Specifically, the first 11 scenarios were used to evaluate the operating characteristics of the proposed designs under the PO assumption, and the remaining 10 scenarios were used to investigate the operating characteristics when the PO assumption was violated.

### 3.2. Simulation results

Empirical power for the PO-based model design under different scenarios is presented in Figure 2. With the calibrated thresholds for futility ($C_f = 0.20$) and superiority ($C_S = 0.96$), we can see from the top plot of Figure 2 that the power increases as the OR increases, given a sample size of 100 for each group during the two stages. The bottom plot of Figure 2 shows that, given the expected OR of 1.8, the sample size increases as the power increases, especially when a sample size of 500 is reached, 83% power is obtained.

[Figure 2]

When the PO assumption violates the calibrated thresholds for futility ($C_f = 0.20$) and superiority ($C_S = 0.95$), the probability of concluding that the treatment is efficacious can be plotted against the sample size for the NPO model–based design (top plot in Figure 3). To evaluate the relations between the ORs and the model's power, we define $OR = (1.0, 1.0, 1.0, 1.0, 1.0)$ as the first scenario (it is a null case), $OR = (1.8, 1.8, 1.8, 1.0, 1.0)$ as the second scenario, and increase it to $OR = (1.8, 1.8, 1.8, 1.7, 1.7)$, with an increment of 0.1 on the last three categories in the OR vector. This setting guarantees a violation of the PO assumption. Under the null hypothesis, the probability is approximately 0.05 and increases with the increasing OR. In the bottom plot of Figure 3, with a predefined OR, for example, $OR = (1.3, 1.5, 1.2, 1, 1.5)$, the power of declaring superiority of the treatment group increases as the sample size increases.

[Figure 3]

Figure 4 displays the performance of the three proposed model-based designs when the assumption of the PO is satisfied. The powers from three model-based designs are computed under different scenarios when the OR varies from 1 to 1.6, with an increment of 0.05. Although all powers increase with the increasing OR, the PO model– and PO/NPO-switch model–based designs perform better than the NPO model–based design, and the discrepancies are obvious when the OR increases. For example, when the OR is 1.6, the power is 77%, 73%, and 73% for the PO, NPO, and PO/NPO-switch model–based designs, respectively. Figure 5 shows the powers for the three proposed model-based designs when the assumption of the PO is not satisfied. Without the assumption, Scenarios 12 to 21 (listed in Table 2) are used. Overall, the three lines show that the power increases when the OR increases. However, with the increasing ORs, the NPO model–based design and PO/NPO-switch model–based design perform better than does the PO model–based design. For instance, in Scenario 20, the empirical powers are 97% by the NPO model–based design, 95% by the PO/NPO-switch model–based design, and 79% by the PO model–based design; the PO model has the lowest power because the assumption is inconsistent with the true data. By comparing the 21 scenarios, we conclude that the PO/NPO-switch model–based design has a robust performance compared to that of the other two designs, regardless of the data-structure assumption. The numeric results are listed in Table 4.

## 4. Implementation in the R package *BayesOrdDesign*

We developed an R package, *BayesOrdDesign*, to implement the proposed methods. The package includes six functions in two chunks. The first chunk includes three functions, ss_ po(·), ss_npo(·), and ss_switch(·), which can be used for estimating the required sample size of the PO, NPO, and PO/NPO-switch model–based designs, respectively. The second chunk has the following functions: get_oc_po(·), get_oc_npo(·), and get_oc_switch(·), which can be used for evaluating the operating characteristics of the three designs, respectively.

Here we use the ss_po(·) function to show how to design a two-stage trial using the PO model–based design as an example and apply the get_oc_npo(·) function to show how to obtain operating characteristics for a two-stage design using the NPO model–based design. We introduce get_oc_switch(·) function, which implements the RJMCMC approach to select the appropriate model-based design and obtain the relevant operating characteristics.

### 4.1. The ss_po function

In this example, we assume that the true proportion of the outcome categories is (0.58, 0.05, 0.17, 0.03, 0.04, and 0.13). Given a clinically meaningful effect size **or_alt**, the function ss_po(·) can estimate the required sample size for the specified type-I/II errors. For example, **alpha** = 0.05 and **power** = 0.8. We ran the simulation 1000 times. Results can be obtained by executing the following codes:

```
result_po = ss_po(or_alt = 2, pro_ctr = c(0.58,0.05,0.17,0.03,0.04,0.13), alpha = 0.05, power = 0.8, nmax = 200, ntrial = 1000, method ="Bayesian")
```

This setup enables the design to search an optimal sample size from 50 to a user-defined upper limitation, which can be defined using the **nmax** argument.

```
result_po
## $total_sample_size_for_each_group
## [1] 100
##
## $power
## [1] 87
##
## $threshold
##   futility superiority
##       0.20       0.99
##
## $typeIerror
## [1] 0.05
```

We can see from the above outputs that the futility and superiority cutoff points (0.20 and 0.99) are obtained to achieve the targeted type-I/II error rates, which in this case are 5% and 13%, respectively, (87% power), with the estimated two-stage sample size of 100 for each group. Thus, the total sample size is 200. Similarly, function ss_npo(·) can be employed to calculate the optimal sample size when the PO assumption is suspicious.

## 4.2. The get_oc_npo function

This example illustrates how to examine the operating characteristics for the NPO model–based design under different scenarios. Unlike the PO model, with only one expected OR input detected in the $H_1$ in the ss_po(·) function, the NPO model–based design needs a vector of ORs for each outcome category as an input parameter in the $H_1$ for the NPO-associated functions. Therefore, the user needs to generate a vector/matrix specifying the alternative ORs. For instance, the first row represents the null hypothesis with all ORs as 1; the second row represents Scenario 2, where $or_1 = or_2 = 1.5$, and $or_3 = or_4 = or_5 = 1$; the third row represents the Scenario 3, where $or_1 = or_2 = 1.5$, and $or_3 = or_4 = or_5 = 1.05$, and so on. The argument **ors** is an 8 by 5 matrix composed of the scenarios in which the user is interested.

```
ors = matrix(c(1.5,1.5,1,1,1,
       1.5,1.5,1.1,1.1,1.1,
       1.5,1.5,1.15,1.15,1.15,
       1.5,1.5,1.2,1.2,1.2,
       1.5,1.5,1.25,1.25,1.25,
       1.5,1.5,1.3,1.3,1.3,
       1.5,1.5,1.35,1.35,1.35), nrow=8, ncol=5, byrow=TRUE)

ors

##      [,1] [,2] [,3] [,4] [,5]
## [1,]  1.5  1.5 1.00 1.00 1.00
## [2,]  1.5  1.5 1.10 1.10 1.10
## [3,]  1.5  1.5 1.15 1.15 1.15
## [4,]  1.5  1.5 1.20 1.20 1.20
## [5,]  1.5  1.5 1.25 1.25 1.25
## [6,]  1.5  1.5 1.30 1.30 1.30
## [7,]  1.5  1.5 1.35 1.35 1.35
```

Then we can input the utility scores (**U**) for each outcome category elicited from the user, and the proportion of each outcome category (**pro_ctr**) for the control based on historical data. In the following code, we show how to examine the operating characteristics if enrolling 200 patients each for the treatment and control groups at each stage, so we specify arguments **fixed_ss** = 200, and **ntrial** = 1,000 indicates the number of simulated trials. Type I error rate **alpha** = 0.05 (under the PO model with OR assumed to be 1.8 and to have 80% power) are specified to calibrate the cutoffs.

```
result_npo = get_oc_npo(alpha = 0.05, pro_ctr = c(0.58,0.05,0.17,0.03,0.04,0.13), U = c(100,80,65,25,10,0), fixed_ss = 200, ors, ntrial = 1000, method = "Bayesian")
```

```
result_npo

## $design
##            Effect Size PET(%) Power(%) Avg SS
## Scenario 1       3.22   0.54   71.35    399
## Scenario 2       4.11   0.11   86.69    400
## Scenario 3       4.51   0.11   91.43    400
## Scenario 4       4.88   0.11   92.75    400
## Scenario 5       5.22   0.00   95.43    400
## Scenario 6       5.54   0.00   98.49    400
## Scenario 7       5.85   0.00   99.03    400
##
## $threshold
##   futility superiority
##      0.20       0.85
##
## $typeIerror
## [1] 0.035
```

The above outputs show the operating characteristics for eight scenarios. Of note, the average sample size (Avg SS) represents the average total sample size for the two stages for each group. Under the null case and given 200 samples per group, we see that the empirical type-I error rate is 4.1%, the futility threshold is 0.20, and the superiority threshold is 0.85. Given the utility scores for each outcome category (100, 80, 65, 25, 10, 0) and the OR vector for Scenario 1 $or_1 = or_2 = 1.5$ and $or_3 = or_4 = or_5 = 1$, the computed equivalent effect size, in terms of the difference in the mean utility scores for the treatment and control, is 3.22. To detect this effect size, given the same sample size under the null case, the empirical power is 71.35%, and the probability of early termination (PET) is 0.54%, with an average sample size of 399 for each group. From the above output, we can also see that with increased effect size, the power increases and the PET decreases, which results in larger total Avg SS. Similarly, the function get_oc_po(·) can be used to evaluate the operating characteristics for the PO model–based design when the PO assumption holds. More details about get_oc_po(·) can be found in the R package *BayesOrdDesign*.

## 4.3. The ss_switch and get_oc_ switch functions

The two core functions of the *BayesOrdDesign* package are ss_switch(·) and get_oc_switch(·), which incorporate R package rjmcmc and allow "jumping" between two candidate models PO and NPO. The function ss_switch(·) can be useful when the clinical team incorrectly decides the structure of the data, such that clinicians believe the data follows a PO assumption, when the true data violates the PO assumption and vice versa. Based on the first-stage data at the interim analysis, the function ss_switch(·) can make an appropriate decision about which model should be used for the rest of the data. Suppose the OR vector of the true data is $or_1 = 1.6, or_2 = or_3 = 1.5, or_4 = or_5 = 1.4$, but clinicians propose an alternative hypothesis with OR = 1.5, because they believe that the data follow a PO assumption (e.g., based on historical data).

As we did with functions ss_po(·) and ss_npo(·), we define the utility function **U** and the distribution of outcome categories **pro_ctr** for the control group and the constraint type-I error rate $\alpha$ and power $\beta$ to calibrate the cutoffs. We also need to provide the number of simulated trials **ntrial** for the null case and alternative cases.

Alternative scenarios can be specified through the argument **or_alt**. Although the PO/NPO-switch model–based design does not require the user to specify the appropriate model, it needs the user to calculate the optimal sample size based on ss_po(·) and ss_npo(·), respectively, before employing the function ss_switch(·) to ensure the power. For instance, to find the optimal sample size under the alternative scenario $or_1 = 1.6, or_2 = or_3 = 1.5, or_4 = or_5 = 1.4$, we execute the ss_npo(·) function to obtain the sample size as below (the argument **nmax** is a user-defined upper limit of the sample size for each group in each stage. If the desirable power cannot be achieved, a larger **nmax** should be tried. A similar rationale applies to the function ss_po(·).

```
results_npo = ss_npo(nmax = 200, or_alt = c(1.6,1.5,1.5,1.4,1.4), pro_ctr = c(0.58,0.05,0.17,0.03,0.04,0.13), U = c(100,80,65,25,10,0), alpha = 0.05, power = 0.8, ntrial = 200, method = "Bayesian")
results_npo
## $total_sample_size_for_each_group
## [1] 75
##
## $power
## [1] 90.51
##
```

```
## $threshold
##   futility superiority
##      0.20       0.95
##
## $typeIerror
## [1] 0.053
```

The readouts show that the calibrated cutoffs are 0.20 and 0.95, and the controlled type-I error rate is 0.053. Under the defined alternative scenario, the optimal two-stage sample size is 75 for the treatment and control groups (150 in total for this trial) to achieve 90.51% power. However, for the PO model–based design, the ORs of the alternative scenario are the same across the outcome categories. Therefore, an alternative OR of 1.5 should be used to find the optimal sample size here. To calculate the corresponding required sample size with the alternative scenarios by ss_po(·) function, we can use the following code:

```
result_po = ss_po(or_alt = 1.5, pro_ctr = c(0.58,0.05,0.17,0.03,0.04,0.13), alpha = 0.05, power = 0.8, nmax = 1000, ntrial = 1000, method ="Bayesian")

result_po

## $total_sample_size_for_each_group
## [1] 475
##
## $power
## [1] 81.00
##
## $threshold
##   futility superiority
##      0.20       0.99
##
## $typeIerror
## [1] 0.045
```

According to the readouts from the PO model–based design, a sample size of 475 for each group during the two stages is required to achieve 81% power; the type-I error rate is controlled at 0.045, and the cutoffs are 0.20 and 0.99, respectively. Now we find the optimal sample sizes of the PO/NPO-switch model–based design for each group and each stage are $n_{po} = 475$ and $n_{npo} = 75$ (from the above functions). In addition, to guarantee that the design achieves the desirable power, the sample size recruited in the second stage, as determined by the proposed design algorithm, will be increased from the initial sample size $n_{po}$ or $n_{npo}$ at incremental steps of 10, until 80% power is achieved. The increased sample size is used when the trial continues to enroll patients during the second stage. Specifically, at interim analysis, the

RJMCMC selects the model-based initial sample size for the sample-size estimation procedure. With all information obtained above, now we can calculate the optimal sample size for the PO/NPO-switch model–based design.

```
result_switch = ss_switch(alpha = 0.05, power = 0.8, n_po = 475, n_npo = 75, n_range = 10, or_alt = c(1.5,1.5,1.5,1.5,1.5), pro_ctr = c(0.58,0.05,0.17,0.03,0.04,0.13), U = c(100,80,65,25,10,0), ntrial = 100, method = "Bayesian")

result_switch

## $total_sample_size_for_each_group
## [1] 577
##
## $power
## [1] 91.30
##
## $threshold
##   futility superiority
##       0.20        0.99
##
## $typeIerror
## [1] 0.03

## $model_selection
##      PO(%) NPO(%)
##       6.67  93.33
```

The calibrated cutoffs are 0.20 and 0.99; they control the type-I/II error rate at 0.03. The optimal two-stage sample size is 577 for both groups. Thus, the total sample size of this trial is 1154, and the power gain is 91.30% in 100 simulated trials. We also present the probability of choosing each model as the best model. Specifically, of the 100 simulated trails, 6.67% selected the PO model–based design at the interim analysis, and 93.33% selected the NPO model–based design.

The function get_oc_switch(·) obtains design parameters, type-I error rate, operating characteristics of the PO/NPO-switch model–based, and two-stage design. We first define a matrix to reflect the scenarios in which we are interested, and each row represents a scenario.

```
ors = matrix(c(1.5,1.5,1.1,1.1,1.1,
               1.5,1.5,1.2,1.2,1.2,
               1.5,1.5,1.3,1.3,1.3), ncol=5, byrow=TRUE)

ors

##      [,1] [,2] [,3] [,4] [,5]
## [1,]  1.5  1.5 1.10 1.10 1.10
```

```
## [2,]  1.5  1.5 1.20 1.20 1.20
## [3,]  1.5  1.5 1.30 1.30 1.30
```

We execute the get_oc_switch(·) function and obtain the readouts shown below. Due to the application of rjmcmc in each simulation, the computational time is much longer than those of the previous functions.

```
result_switch = get_oc_Switch(alpha = 0.05, pro_ctr = c(0.58,0.05,0.17,0.03,0.04,0.13), U = c(100,80,65,
25,10,0), ors, n_po = 100, n_npo = 200, ntrial = 1000, method = "Bayesian")

result_switch

## $design
##            Effect Size  PET(%)  Power(%)  Avg SS   PO(%)   NPO(%)
## Scenario 1    4.11       0.00    91.33    563.33    3.33    96.67
## Scenario 2    4.88       3.57    92.87    561.61    3.77   100.00
## Scenario 3    5.54       0.00    93.33    550.00    0.00   100.00
##
## $threshold
##   futility superiority
##      0.20        0.85
##
## $typeIerror
## [1] 0.041
```

## 5. Conclusion

Although ordinal endpoints as the primary outcomes have been used in various clinical trials, discussions from the trial-design perspective have been limited. Among the existing methods, most are based the PO assumption, and few were developed for the NPO assumption. Selecting the appropriate model is a practical question for clinical trials employing ordinal endpoints. For the first time, in this paper, we provide novel Bayesian two-stage designs for the ordinal endpoints, which include PO model– and NPO model–based designs, and a robust PO/NPO-switch model–based design, if there are uncertainties about which assumption (PO or NPO) is appropriate before the trial is initiated.

Simulation results demonstrated that the PO/NPO-switch model is robust, regardless of the model's assumptions. Specifically, when the PO assumption is satisfied, the PO model–based design has the highest probability of recommending the efficacious treatment group. When the PO assumption is violated, then the NPO model–based design shows superior performance.

However, the empirical powers of the PO/NPO-switch model, under various scenarios, are higher than the PO model–based design when the PO assumption is violated, or higher than the NPO model–based design when the PO assumption is met.

In summary, the proposed PO/NPO-switch model is a robust, midway solution compared to the other two designs. For practitioners to implement these methods in an accessible way, we have also developed an R package, *BayesOrdDesign*, to compute the required sample size and power for the three different model designs. For the PO/NPO-switch model–based design, we show how to switch the model based on the interim data, which has the potential to save sample size. However, the sample size re-estimation approach is ad hoc in this paper. We may employ a more rigorous way using the posterior predictive probability to re-estimate the sample size from the interim data, which will be our future project.

## 6. References

Barker, R. J., & Link, W. A. (2013). Bayesian multimodel inference by RJMCMC: A Gibbs sampling approach. *American Statistician*, *67*(3), 150–156. https://doi.org/10.1080/00031305.2013.791644

Bath, P. M. W., Gray, L. J., Collier, T., Pocock, S., & Carpenter, J. (2007). Can we improve the statistical analysis of stroke trials? Statistical reanalysis of functional outcomes in stroke trials. *Stroke*, *38*(6), 1911–1915. https://doi.org/10.1161/STROKEAHA.106.474080

Choi, S. Y. (2016). *Statistical Considerations for the Analysis of Ordinal Outcomes in Randomized Controlled Trials*.

Christensen RHB. (2019). *"ordinal—Regression Models for Ordinal Data ." R package version 2019.12-10. https://CRAN.R-project.org/package=ordinal*.

D'Amico, G., Abraldes, J. G., Rebora, P., Valsecchi, M. G., & Garcia-Tsao, G. (2020). Ordinal Outcomes Are Superior to Binary Outcomes for Designing and Evaluating Clinical Trials in Compensated Cirrhosis. *Hepatology*, *72*(3), 1029–1042. https://doi.org/10.1002/hep.31070

Desantis, S. M., Lazaridis, C., Palesch, Y., & Ramakrishnan, V. (2014). Regression analysis of ordinal stroke clinical trial outcomes: An application to the NINDS t-PA trial. *International Journal of Stroke*, *9*(2), 226–231. https://doi.org/10.1111/ijs.12052

DeSantis, S. M., Lazaridis, C., Palesch, Y., & Ramakrishnan, V. (2014). Regression Analysis of Ordinal Stroke Clinical Trial Outcomes: An Application to the NINDS t-PA Trial. *International Journal of Stroke*, *9*(2), 226–231. https://doi.org/10.1111/ijs.12052

Diener, H. C., Lees, K. R., Lyden, P., Grotta, J., Davalos, A., Davis, S. M., Shuaib, A., Ashwood, T., Wasiewski, W., Alderfer, V., Hårdemark, H. G., & Rodichok, L. (2008). NXY-059 for the treatment of acute stroke: Pooled analysis of the SAINT I and II trials. *Stroke*, *39*(6), 1751–1758. https://doi.org/10.1161/STROKEAHA.107.503334

Gelling, N., Schofield, M. R., & Barker, R. J. (2018). *R package rjmcmc : The calculation of posterior model probabilities from MCMC output Transdimensional algorithms*. 1–28.

Gordon D. Murray, David Barer, Sung Choi, Helen Fernandes, Barbara Gregson, Kennedy R. Lees, Andrew I.R. Maas, Anthony Marmarou, A. David Mendelow, Ewout W. Steyerberg, Gillian S. Taylor, Graham M. Teasdale, and C. J. Weir. (2005). Design and Analysis of Phase III Trials with Ordered Outcome Scales: The Concept of the Sliding Dichotomy. *Journal of Neurotrauma.*, 511-517. https://doi.org/http://doi.org/10.1089/neu.2005.22.511

Green, P. J. (1995). Reversible Jump Markov Chain Monte Carlo Computation and Bayesian Model Determination. *Biometrika*, *82*(4), 711. https://doi.org/10.2307/2337340

Guo, B., & Yuan, Y. (2017). Bayesian Phase I/II Biomarker-Based Dose Finding for Precision Medicine With Molecularly Targeted Agents. *Journal of the American Statistical Association*, *112*(518), 508–520. https://doi.org/10.1080/01621459.2016.1228534

Harrell, F., & Lindsell, C. (2021). *Statistical Design and Analysis Plan for Sequential Parallel-Group RCT for COVID-19*. http://hbiostat.org/proj/covid19/bayesplan.html

Houston, B. L., Lawler, P. R., Goligher, E. C., Farkouh, M. E., Bradbury, C., Carrier, M., Dzavik, V., Fergusson, D. A., Fowler, R. A., Galanaud, J. P., Gross, P. L., McDonald, E. G., Husain, M., Kahn, S. R., Kumar, A., Marshall, J., Murthy, S., Slutsky, A. S., Turgeon, A. F., … Zarychanski, R. (2020). Anti-Thrombotic Therapy to Ameliorate Complications of COVID-19 (ATTACC): Study design and methodology for an international, adaptive Bayesian randomized controlled trial. *Clinical Trials*, *17*(5), 491–500. https://doi.org/10.1177/1740774520943846

Howard, G., Waller, J. L., Voeks, J. H., Howard, V. J., Jauch, E. C., Lees, K. R., Nichols, F. T., Rahlfs, V. W., & Hess, D. C. (2012). A simple, assumption-free, and clinically interpretable approach for analysis of modified Rankin outcomes. *Stroke*, *43*(3), 664–669. https://doi.org/10.1161/STROKEAHA.111.632935

Liberman, L., & Menell, J. H. (2002). Breast imaging reporting and data system (BI-RADS). *Radiologic Clinics of North America*, *40*(3), 409–430, v. https://doi.org/10.1016/s0033-8389(01)00017-3

McCullagh, P. (1980). Regression Models for Ordinal Data. *Journal of the Royal Statistical Society: Series B (Methodological)*, *42*(2), 109–127. https://doi.org/10.1111/j.2517-6161.1980.tb01109.x

Murray, T. A., Thall, P. F., & Yuan, Y. (2016). Utility-based designs for randomized comparative trials with categorical outcomes. *Statistics in Medicine*, *35*(24), 4285–4305. https://doi.org/10.1002/sim.6989

Murray, T. A., Yuan, Y., Thall, P. F., Elizondo, J. H., & Hofstetter, W. L. (2018). A utility-based design for randomized comparative trials with ordinal outcomes and prognostic subgroups. *Biometrics*, *74*(3), 1095–1103. https://doi.org/10.1111/biom.12842

Peterson, R. L., Vock, D. M., Babiker, A., Powers, J. H., Hunsberger, S., Angus, B., Paez, A., & Neaton, J. D. (2019). Comparison of an ordinal endpoint to time-to-event, longitudinal, and binary endpoints for use in evaluating treatments for severe influenza requiring hospitalization. *Contemporary Clinical Trials Communications*, *15*(May), 100401. https://doi.org/10.1016/j.conctc.2019.100401

Roozenbeek, B., Lingsma, H. F., Perel, P., Edwards, P., Roberts, I., Murray, G. D., Maas, A. I. R., & Steyerberg, E. W. (2011). The added value of ordinal analysis in clinical trials: An example in traumatic brain injury. *Critical Care*, *15*(3). https://doi.org/10.1186/cc10240

Sakamaki, K., Uemura, Y., & Shimizu, Y. (2021). Definitions and elements of endpoints in phase III randomized trials for the treatment of COVID-19: a cross-sectional analysis of trials registered in ClinicalTrials.gov. *Trials*, *22*(1), 1–9. https://doi.org/10.1186/s13063-021-05763-y

Synopsis, C.-T. T. (2020). Novel Coronavirus WHO COVID-19 Therapeutic Trial Synopsis, February 18, 2020, Geneva, Switzerland. *Situation Report – 205*, *205*(6), 1–19.

Whitehead, J. (1993). Sample size calculations for ordered categorical data. *Statistics in Medicine*, *12*(24), 2257–2271. https://doi.org/10.1002/sim.4780122404

Whitehead, J., & Horby, P. (2017). GOST: A generic ordinal sequential trial design for a treatment trial in an emerging pandemic. *PLoS Neglected Tropical Diseases*, *11*(3), 1–13. https://doi.org/10.1371/journal.pntd.0005439

Whitehead, J., & Jaki, T. (2009). One- and two-stage design proposals for a phase II trial comparing three active treatments with control using an ordered categorical endpoint. *Statistics in Medicine*, *28*(5), 828–847. https://doi.org/10.1002/sim.3508

**Table 1.** Ordered categorical data available at the end of a clinical trial.

| Groups | 1 | 2 | ... | $C$ | Total |
|---|---|---|---|---|---|
| Control | $n_{11}$ | $n_{12}$ | ... | $n_{1C}$ | $n_{1\cdot}$ |
| Experimental | $n_{21}$ | $n_{22}$ | ... | $n_{2C}$ | $n_{2\cdot}$ |
| Total | $n_{\cdot 1}$ | $n_{\cdot 2}$ | ... | $n_{\cdot C}$ | $n$ |

**Table 2.** Scenarios to be investigated in the simulation study. The proportions of the control group $q_{jc}$ are $(0.58, 0.05, 0.17, 0.03, 0.04, 0.13)$ for all scenarios.

| Model-based | Scenario | $\Delta$ | $p_{ic}$ | | | | | | $\bar{U}_c$ | $\bar{U}_t$ | $\bar{U}_t - \bar{U}_c$ |
|---|---|---|---|---|---|---|---|---|---|---|---|
| PO | 1 | 1.10 | 0.6 | 0.05 | 0.16 | 0.03 | 0.04 | 0.12 | 74.20 | 75.88 | 1.68 |
| | 2 | 1.15 | 0.61 | 0.05 | 0.16 | 0.03 | 0.04 | 0.11 | | 76.64 | 2.44 |
| | 3 | 1.20 | 0.62 | 0.05 | 0.16 | 0.03 | 0.04 | 0.11 | | 77.35 | 3.15 |
| | 4 | 1.25 | 0.63 | 0.05 | 0.15 | 0.03 | 0.03 | 0.11 | | 78.02 | 3.82 |
| | 5 | 1.30 | 0.64 | 0.05 | 0.15 | 0.03 | 0.03 | 0.1 | | 78.65 | 4.45 |
| | 6 | 1.35 | 0.65 | 0.05 | 0.15 | 0.02 | 0.03 | 0.1 | | 79.25 | 5.05 |
| | 7 | 1.40 | 0.66 | 0.05 | 0.14 | 0.02 | 0.03 | 0.1 | | 79.81 | 5.61 |
| | 8 | 1.45 | 0.67 | 0.04 | 0.14 | 0.02 | 0.03 | 0.09 | | 80.34 | 6.14 |
| | 9 | 1.50 | 0.67 | 0.04 | 0.14 | 0.02 | 0.03 | 0.09 | | 80.85 | 6.65 |
| | 10 | 1.55 | 0.68 | 0.04 | 0.14 | 0.02 | 0.03 | 0.09 | | 81.32 | 7.12 |
| | 11 | 1.60 | 0.69 | 0.04 | 0.13 | 0.02 | 0.03 | 0.09 | | 81.78 | 7.58 |
| NPO | 12 | (1.2, 1.2, 1, 1, 1) | 0.62 | 0.05 | 0.13 | 0.03 | 0.04 | 0.13 | | 75.69 | 1.49 |
| | 13 | (1.3, 1.2, 1.1, 1, 1) | 0.64 | 0.03 | 0.14 | 0.02 | 0.04 | 0.13 | | 76.66 | 2.46 |
| | 14 | (1.4, 1.3, 1.1, 1.1, 1) | 0.66 | 0.03 | 0.13 | 0.03 | 0.03 | 0.13 | | 77.45 | 3.25 |
| | 15 | (1.4, 1.3, 1.2, 1.1, 1) | 0.66 | 0.03 | 0.14 | 0.02 | 0.03 | 0.13 | | 77.96 | 3.76 |
| | 16 | (1.4, 1.3, 1.3, 1.2, 1) | 0.66 | 0.03 | 0.15 | 0.02 | 0.02 | 0.13 | | 78.58 | 4.38 |
| | 17 | (1.4, 1.4, 1.4, 1.2, 1.1) | 0.66 | 0.05 | 0.14 | 0.01 | 0.03 | 0.12 | | 79.31 | 5.11 |
| | 18 | (1.5, 1.5, 1.4, 1.2, 1.1) | 0.67 | 0.04 | 0.13 | 0.01 | 0.03 | 0.12 | | 79.82 | 5.62 |
| | 19 | (1.6, 1.6, 1.4, 1.3, 1) | 0.69 | 0.04 | 0.12 | 0.02 | 0.01 | 0.13 | | 80.34 | 6.14 |

| | 20 | (1.7, 1.6, 1.5, 1.3, 1) | 0.7 | 0.03 | 0.13 | 0.01 | 0.01 | 0.13 | | 80.94 | 6.74 |
| | 21 | (1.8, 1.6, 1.5, 1.4, 1) | 0.71 | 0.02 | 0.13 | 0.02 | 0 | 0.13 | | 81.31 | 7.11 |

**Abbreviations:** NPO, nonproportional odds; PO, proportional odds

**Table 3.** Empirical type-I error rates of three designs and the corresponding thresholds.

| Design | $C_f$ | $C_s$ | Type-I Error |
|---|---|---|---|
| PO model | 0.20 | 0.96 | 0.046 |
| NPO model | 0.20 | 0.95 | 0.052 |
| PO/NPO-switch model | 0.20 | 0.99 | 0.052 |

**Abbreviations:** NPO, nonproportional odds; PO, proportional odds

**Table 4.** Empirical power, probability of early termination, and total number of average sample sizes of patients under different scenarios adopting the three proposed designs.

| Proposed design | Scenario | PET (%) | | | Power (%) | | | Avg SS | | |
|---|---|---|---|---|---|---|---|---|---|---|
| | | PO | NPO | Switch | PO | NPO | Switch | PO | NPO | Switch |
| **Model assumption** | | | | | | | | | | |
| PO | 1 | 0.05 | 0.05 | 0.11 | 0.13 | 0.13 | 0.14 | 391 | 391 | 379 |
| | 2 | 0.04 | 0.03 | 0.09 | 0.19 | 0.16 | 0.17 | 393 | 395 | 381 |
| | 3 | 0.02 | 0.02 | 0.08 | 0.26 | 0.19 | 0.23 | 396 | 397 | 385 |
| | 4 | 0.01 | 0.01 | 0.04 | 0.33 | 0.27 | 0.28 | 398 | 399 | 391 |
| | 5 | 0.01 | 0 | 0.04 | 0.40 | 0.32 | 0.35 | 399 | 400 | 392 |
| | 6 | 0 | 0.01 | 0.03 | 0.48 | 0.39 | 0.44 | 399 | 399 | 394 |
| | 7 | 0 | 0 | 0.02 | 0.54 | 0.50 | 0.50 | 400 | 400 | 396 |
| | 8 | 0 | 0 | 0.01 | 0.62 | 0.56 | 0.57 | 400 | 400 | 397 |
| | 9 | 0 | 0 | 0.01 | 0.68 | 0.62 | 0.60 | 400 | 400 | 398 |
| | 10 | 0 | 0 | 0 | 0.73 | 0.69 | 0.67 | 400 | 400 | 399 |
| | 11 | 0 | 0 | 0.01 | 0.77 | 0.73 | 0.73 | 400 | 400 | 399 |
| NPO | 12 | 0 | 0.06 | 0 | 0.05 | 0.13 | 0.01 | 399 | 389 | 400 |
| | 13 | 0 | 0.01 | 0 | 0.10 | 0.24 | 0.03 | 400 | 398 | 400 |
| | 14 | 0 | 0 | 0 | 0.21 | 0.43 | 0.10 | 400 | 399 | 400 |
| | 15 | 0 | 0 | 0 | 0.23 | 0.51 | 0.18 | 400 | 400 | 400 |
| | 16 | 0 | 0 | 0 | 0.25 | 0.66 | 0.31 | 400 | 399 | 400 |
| | 17 | 0 | 0 | 0 | 0.36 | 0.80 | 0.48 | 400 | 400 | 400 |

| | | | | | | | | | |
|---|---|---|---|---|---|---|---|---|---|
| | 18 | 0 | 0 | 0 | 0.54 | 0.88 | 0.72 | 400 | 400 | 400 |
| | 19 | 0 | 0 | 0 | 0.67 | 0.94 | 0.87 | 400 | 400 | 400 |
| | 20 | 0 | 0 | 0 | 0.79 | 0.97 | 0.95 | 400 | 400 | 400 |
| | 21 | 0 | 0 | 0 | 0.86 | 0.98 | 0.99 | 400 | 400 | 400 |

**Abbreviations:** Avg SS, average sample size; NPO, nonproportional odds; PET, probability of early termination; PO, proportional odds

**Figure 1.** PO/NPO-switch model–based design

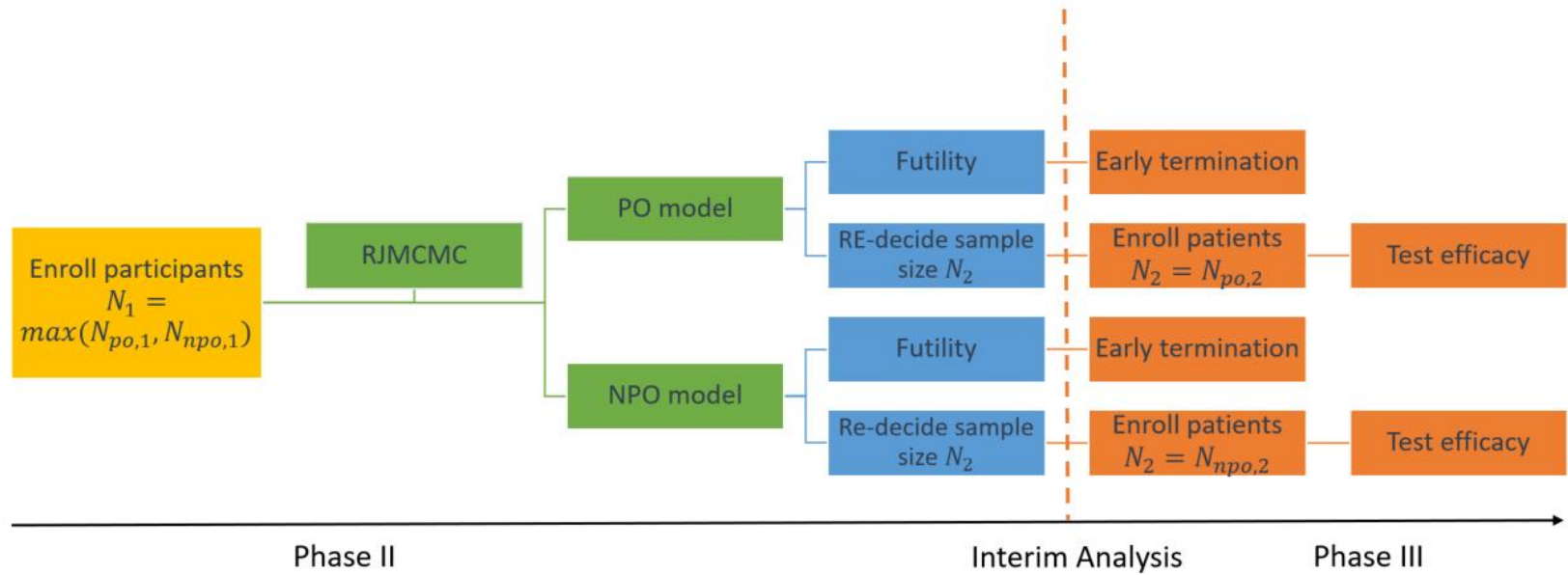

**Figure 2.** Empirical power of the PO model–based design for different true-effect sizes between the treatment and control groups (top) and for different sample sizes per group per stage (bottom).

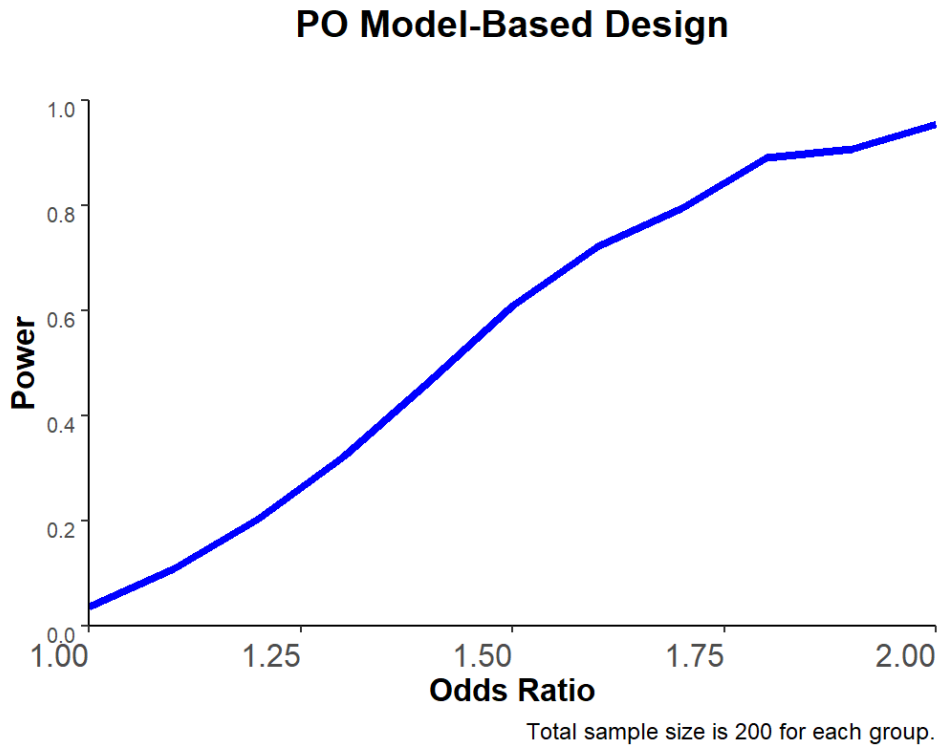

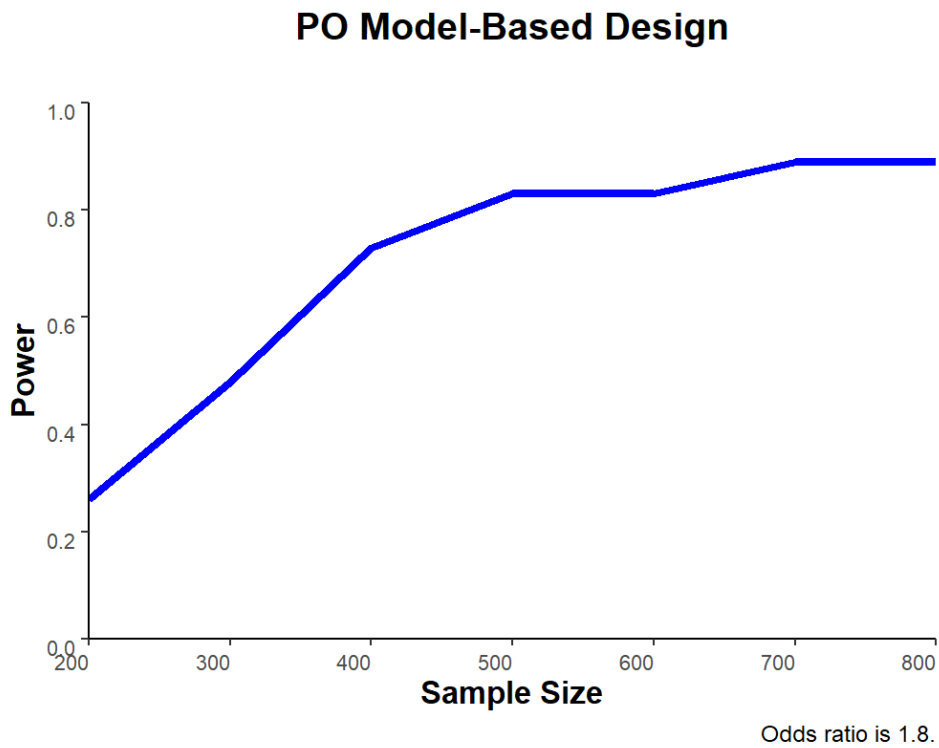

**Figure 3.** Empirical power of the NPO model–based design for different true-effect sizes between the treatment and control groups (top) and for different sample sizes per group per stage (bottom).

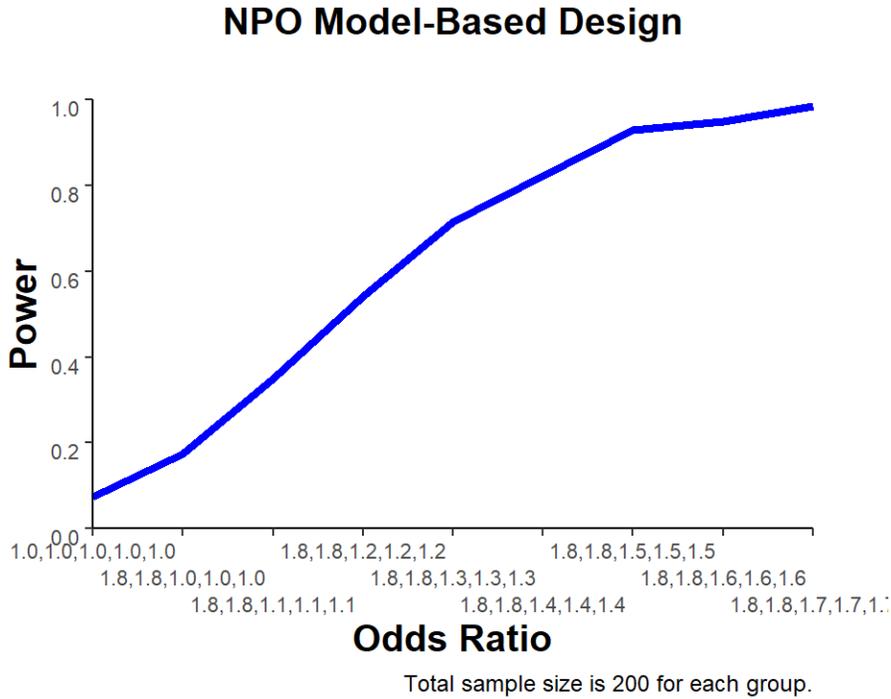

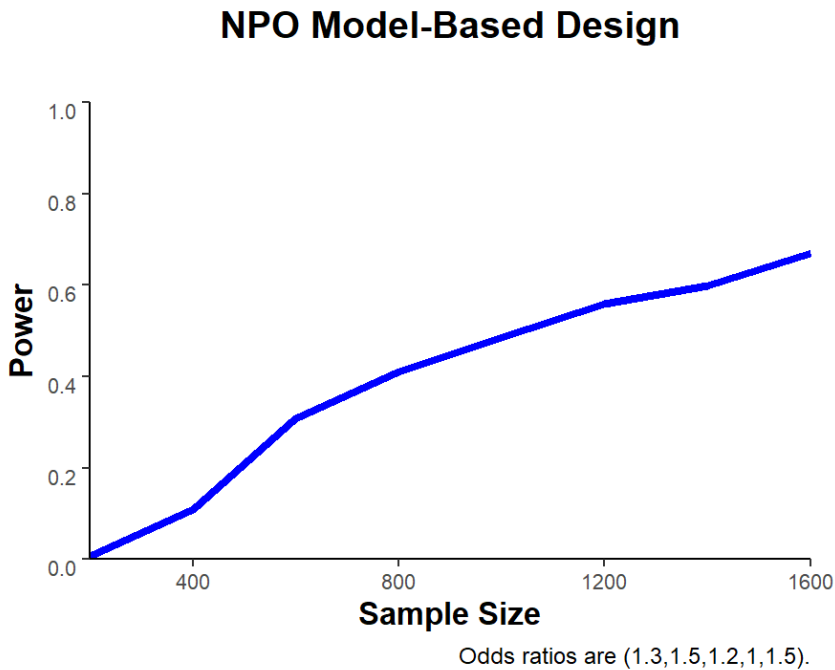

**Figure 4.** Comparison of empirical power of the PO (blue), NPO (yellow), and PO/NPO-switch (red) model–based designs for different true-effect sizes between the treatment and control groups when the PO assumption is satisfied. Sample size for each group = 200.

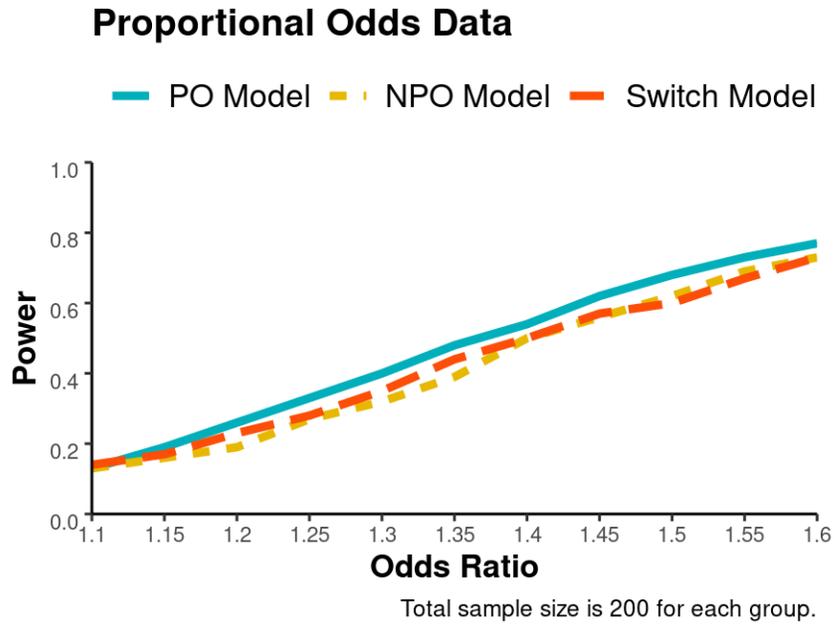

**Figure 5.** Comparison of the empirical power of the PO (blue), NPO (yellow), and PO/NPO-switch model–based designs for different true-effect sizes between the treatment and control groups when the PO assumption violates. The sample size for each group = 200.

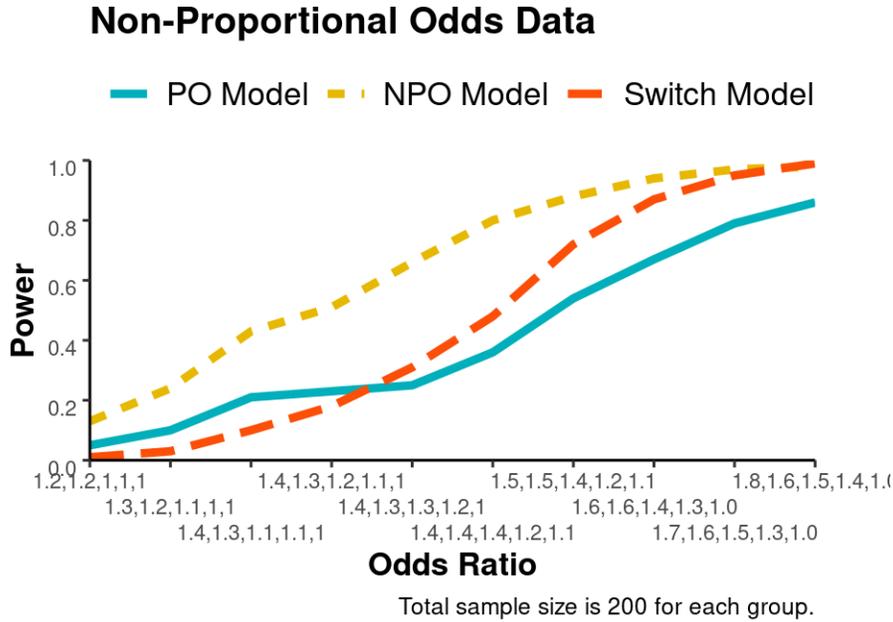

# Appendix

## 1. Priors on parameters

We specify the prior distribution for μ, γ, θ, and Δ as follows:

$$\mu \sim Normal(0,1),$$

$$\gamma_c \sim Normal(0, 0.1), c = 1, \ldots, 5$$

$$\theta_c \sim Normal(0, 0.1), c = 1, \ldots, 5$$

$$\Delta_c \sim Normal(\widehat{\Delta}_c, \hat{\sigma}_{\Delta_c}^2)$$

For the PO model, γ and θ are identical and only one Δ is required. Assume that the outcome follows a categorical distribution, which is a discrete probability distribution that describes possible results of a random variable that can take on one of six possible categories, with probability of each category obtained separately.

$$x, y | \mu, \gamma, \theta, \Delta \sim \text{Categorical}(p_{i,j}), i = 1, \ldots, N; j = 1, \ldots, 6$$

Where $p_c$ represents the probability of corresponding categorical value, and $\sum_{c=1}^{6} p_c = 1$.

## 2. RJMCMC

For the PO model, the likelihood function for a single observation is

$$p(x, y = c | \mu, \gamma, \Delta, M_1) \propto \frac{1}{1 + e^{-\gamma_c + \mu + \Delta * x}} - \frac{1}{1 + e^{-\gamma_{c-1} + \mu + \Delta * x}}.$$

Here, $x \in (0,1)$, $x = 0$ denotes that the observation is in the control group and vice versa. The $y$ denotes the responses for each observation. Given the equality between γ and θ, γ is employed to represent the latent thresholds for both groups. For a sample of $n$ independent and identically distributed observations, $x_1, \ldots, x_n, y_1, \ldots, y_n$, is

$$p(x_1, \ldots, x_n, y_1, \ldots, y_n | \mu, \gamma, \Delta, M_1) = \prod_{i=1}^{n} p(x_i, y_i | \mu, \gamma, \Delta, M_1).$$

For the NPO model, the likelihood function for a single observation is

$$p(x = 0, y = c | \mu, \gamma, \Delta_c, M_2) \propto \frac{1}{1 + e^{-\gamma_c + \mu}} - \frac{1}{1 + e^{-\gamma_{c-1} + \mu}}, \text{ and}$$

$$p(x = 1, y = c | \mu, \gamma, \Delta_c, M_2) \propto \frac{1}{1 + e^{-\theta_c + \mu + \Delta_c}} - \frac{1}{1 + e^{-\theta_{c-1} + \mu + \Delta_c}}.1$$

The likelihood for all observations is

$$p(x_1, \ldots, x_n, y_1, \ldots, y_n | \mu, \gamma, \theta, \Delta, M_2) = \prod_{i=1}^{n} p(x_i, y_i | \mu, \gamma, \theta, \Delta, M_2).$$

Then, prior distributions for $\psi$ are calculated as shown below.

$$p(\psi | M_1) = p(\beta_1 | M_1) \times |J_1| = \prod_{c=1}^{5} p(\Delta_c | M_1) \times 1$$

$$p(\psi | M_2) = p(\beta_2 | M_1) \times |J_2| = p(\Delta | M_2) \times \prod_{c=2}^{4} p(\mu_c | M_2) \times |J_2|$$

Determinants of the Jacobian $|J_1|$ and $|J_2|$ are required for the change of variables theorem. Note that the determinants do not require manual calculation, as the package automatically calculates them. Due to the irrelevance between the supplemental variable vector $\boldsymbol{\mu}$ and $M_2$, the specific choice of $[\boldsymbol{\mu} | M_2]$ has no bearing on inference. Therefore, in our calculation, prior distributions of $\boldsymbol{\mu}$ are ignored to improve the performance of the RJMCMC algorithm.